\documentstyle[12pt]{article}
\begin{document}

\hoffset = -1truecm
\voffset = -2truecm

\begin{flushright}
{KEK-TH-391}\\
{KEK Preprint 93-217}\\
{March 1994}\\
H
\end{flushright}
\vspace*{5mm}

\begin{center}
{\Large \bf
A scalar gluonium contribution to
$K\rightarrow\pi\pi$ decay}

\vspace*{2.0cm}

{\large \bf A.A.Penin}\\
{\it Institute for Nuclear  Research of the Russian Academy of
Sciences,\\
Moscow 117312, Russia}\\
and\\
{\large \bf A.A. Pivovarov
\footnote{On leave from
Institute for Nuclear Research, Moscow, Russia}}\\
{\it National Laboratory for High Energy Physics (KEK),\\
Tsukuba, Ibaraki 305, Japan}\\
\vspace*{1.0cm}
{\large \bf Abstract}\\
\end{center}
\noindent

We study a new $K\rightarrow\pi\pi$ decay channel with
gluons in intermediate state which is
normally neglected within the factorization framework.
Both short-distance and long-distance parts of the
amplitude are calculated.
The chiral Lagrangian approach is used for obtaining
the long-distance contribution. The nonperturbative contribution
gives an additional enhancement to
$K\rightarrow\pi\pi$ decay amplitude with $\Delta I=1/2$.
A sizable violation of factorization
in the $p^4$ order of chiral perturbation theory is found.

\vspace{0.5in}
\noindent
PACS number(s): 12.15.Ji, 13.25.+m, 11.50.Li, 11.30.Rd.

\thispagestyle{empty}
\newpage

The origin of considerable enhancement of $\Delta I=1/2$ parts of
non-leptonic kaon decay amplitudes remains
one of subtle points within the standard model (SM) [1].
Though explained qualitatively by strong interaction effects
it still escapes the reliable quantitative description.
Numerous attempts have been recently made
to achieve a sufficient accuracy for non-leptonic kaon decays in SM.
The efforts were directed
to improving the perturbative QCD analysis
[2] and to accounting for long-distance effects using more
advanced models of strong interactions at low energy,
for example, the chiral  Lagrangians [3], $1/N_c$
expansion of QCD [4,5] or lattice simulations [6]. The present
results however do not fit experimental data.

An effective $\Delta S=1$ Hamiltonian reads [7-9]
$$
H_{\Delta S=1}= {G_F\over\sqrt
2}V_{ud}V^*_{us}\sum^6_{i=1,i\neq
4}[z_i(\mu)+\tau y_i(\mu)]Q_i
\eqno(1)
$$
where $G_F$ is a Fermi constant, $V$ is
the quark flavor mixing matrix,
$\tau=-{V_{td}V^*_{ts}/V_{ud}V^*_{us}}$,
$z_i(\mu)$ and $y_i(\mu)$ are the coefficients of
Wilson expansion, $\{Q_i\}$ is a full basis of dimension six
local operators
containing light quark fields $(u,d,s)$ only.

The renormalization group improved
perturbation theory does not account
for strong interaction of soft
light quarks and gluons. The information
about this interaction is entirely contained in
the hadronic matrix elements of local
quark operators. Factorization
procedure for evaluation of these matrix elements [10],
{\it i.e.} the procedure of replacing the
four-quark operators by a product of two
non-interacting quark currents accounts only
for the "factorizable" part of this interaction [11].
But there are also "non-factorizable"
contributions, for example, ones corresponding to
annihilation of a quark pair from the
four-quark operator into soft gluons, which are
omitted within the factorization procedure.
The calculation of these
contributions and the generalization of
matrix element estimates beyond the factorization
framework can be systematically done within
QCD sum rules technique combined with the
chiral Lagrangian approach.
This possibility is related to studying
a new $K\rightarrow\pi\pi$ decay channel
generated by annihilation of a quark
pair from the four-quark operator into gluons
with the subsequent formation of the final pion pair by
the soft gluon cloud, {\it i.e.} the decay
channel with gluons playing the role of
an intermediate state. Being non-factorizable,
this decay mode does not appear as
a correction to some leading order
contribution and can be studied by its own.
This feature makes obtained results more accurate.

In the present paper we study a new
$K\rightarrow\pi\pi$ decay channel with the
simplest scalar colorless gluon configuration
forming an intermediate state.  We calculate both
short-distance (perturbative) and long-distance
(nonperturbative) parts of the corresponding
amplitude. To obtain the
long-distance contribution, the chiral effective
Lagrangian is used as a low energy model of strong
interactions and the relevant factor of proportionality
between quark and mesonic operators is
derived via QCD sum rules.

Non-factorizable contributions
reveal themselves in two different ways:
first, they appear as corrections to the
couplings characterizing the "factorizable"
weak chiral Lagrangian in $O(p^4)$ and higher orders,
second, some new non-factorizable terms emerge.
The latter is the case for
$K\rightarrow\pi\pi$ decay mode with gluons
forming an intermediate state.

Before turning to the long-distance effects of
meson-gluon transitions it is useful to
consider the similar phenomenon arising already
in perturbative QCD as a leading correction
in the inverse mass of charmed quark.
The effective low energy tree-level Hamiltonian
for $\Delta  S=1$ transitions before decoupling
of the $c$-quark reads
$$
H_{\Delta S=1}= {G_F\over\sqrt
2}V_{ud}V^*_{us}(Q_2^u-(1-\tau )Q_2^c)
\eqno(2)
$$
where $Q_2^q=4(\bar s_L\gamma_\mu
q_L)(\bar q_L\gamma_\mu d_L)$, $q=u, c$ and
$q_{L(R)}$ stands for the left(right) handed quark.
Performing the {\it OPE}
to the first order terms in $\alpha_s$ and
$m_c^{-1}$ one can write down a representation for
the effective Hamiltonian in the form
$$
H_{\Delta S=1}=H_0+H_1.
\eqno(3)
$$
The first addendum $H_0$ in the {\it rhs} of
eq.~(3) corresponds to the leading
contributions in $m_c^{-1}$ and  coincides with
the {\it rhs} of eq.~(1) while
the second addendum $H_1$ is the leading
order $m_c^{-1}$ correction that comes from
an annihilation diagram with the charmed quark
running inside the loop.
If one restricts the analysis to a scalar
colorless gluon configuration
$G^a_{\mu\nu}G^a_{\mu\nu}$ the
additional contribution reads [12]
$$
H_1=-{G_F\over\sqrt 2}V_{ud}V^*_{us}(1-\tau)
{1\over 120 m_c^2}m_s\bar
s_Rd_L{\alpha_s\over\pi}G^a_{\mu\nu}G^a_{\mu\nu}.
\eqno(8)
$$
To derive an effective Lagrangian
involving Goldstone bosons only,
which can be used for the calculation of decay
amplitudes, one has to replace a QCD operator
$$
m_s\bar s_Rd_L{\alpha_s\over\pi}
G^a_{\mu\nu}G^a_{\mu\nu}\equiv m_s\bar s_Rd_L G^2
\eqno(9)
$$
in eq.~(8) with its mesonic counterpart.
The following representation holds
$$
m_s\bar s_Rd_L G^2
=Af_\pi^6(U^\dagger\chi)_{23}+Bf^4_\pi (U^\dagger\chi)_{23}
tr(\partial_\mu U^\dagger\partial^\mu U)
+{\rm other}~O(p^4)~{\rm terms}
\eqno(10)
$$
where  $\chi$ stands for the meson
mass matrix, $U=e^{-i{\sqrt{2}\phi/f_\pi}}$ is
the unitary matrix describing the octet of
pseudoscalar mesons, $A$ and $B$ are
dimensionless parameters to be computed.
The second term on the {\it rhs} of eq.~(10) is
separated from other $O(p^4)$ structures for
the reason that will be clarified below.

The $O(p^2)$ term in eq.~(10) is
a tadpole. The appearance of such
a term is the consequence of working with a
wrong solution for the ground state. This term merely
renormalizes the effective Lagrangian of strong interactions
and can be absorbed into the meson mass matrix by a
suitable $SU(3)_L\otimes SU(3)_R$ rotation [3,14];
it does not affect any observables.
A contribution to the physical amplitude is
determined by the $O(p^4)$ part of the chiral
representation (10). The most transparent way
to obtain this contribution is to consider the
quark-gluon operator (9) as a product of the
(pseudo)scalar quark current and the scalar colorless
gluon operator. Then one can replace the
quark operator by its mesonic realization
according to the {\it PCAC} hypothesis [13]
$$
m_s\bar s_Rd_L\rightarrow
-{f^2_\pi\over 8} (U^\dagger\chi )_{23}.
\eqno(11)
$$
On the other hand there is a low energy theorem
based on fundamental properties of the
energy-momentum tensor that gives the
chiral representation of the gluon operator [15]
$$
{\alpha_s\over\pi}G^a_{\mu\nu}G^a_{\mu\nu}=
-{2\over b}f^2_\pi tr(\partial_\mu U^\dagger
\partial^\mu U)+O(p^4)
\eqno(12)
$$
where $b=9$ for three light quark flavors.
Eqs.~(11,12) give for the $B$ parameter
$$
B={1\over 4b}.
\eqno(13)
$$
This approximation
corresponds to the simplest physical picture
where the kaon is annihilated by the
pseudoscalar quark current while the
pion pair is born by the gluon operator.
Eqs.~(8,10-13) lead to an effective
chiral Lagrangian of the form
$$
L^G_{eff}={G_F\over\sqrt 2}V_{ud}V^*_{us}(1-\tau)
{f^4_\pi\over 480b m^2_c}(U^\dagger \chi)_{23}
tr(\partial_\mu U^\dagger\partial^\mu U).
\eqno(14)
$$
Since the pion pair is born by the gluon operator
this Lagrangian describes the decay
channel with gluons forming an intermediate state.

Now the corresponding $K\rightarrow\pi\pi$  decay
amplitude becomes explicitly calculable. We
use the standard parametrization of the amplitude
$A_0$ with the isospin transfer
$\Delta I=1/2$
$$
ReA_0=
{G_F\over\sqrt 2}\sin\theta_c
\cos\theta_c f_K m^2_K g_{1/2}
\eqno(15)
$$
where $\theta_c$ stands for Cabibbo angle
and $g_{1/2}$ is a dimensionless parameter.
The new contribution reads
$$
\Delta g_{1/2}={m^2_K\over 30b m^2_c}
\sim 10^{-3}
\eqno(16)
$$
while experiment gives [16]
$$
g^{exp}_{1/2}=3.9,
\eqno(17a)
$$
and the most recent theoretical estimate is [2]
$$
g_{1/2}\sim 2.6.
\eqno(17b)
$$
Thus the local (perturbative) part of new decay mode
is negligible according to the general estimate of
a scale of the leading order charmed quark mass
corrections [12].

The local effective Hamiltonian (8)
does not exhaust the whole physics of the
meson-gluon transitions. It can not account for the
long-distance contribution connected with the
propagation of a soft $u$-quark round the loop
of the annihilation diagram. Because of the
lightness of the $u$-quark this contribution can
not be represented as a local vertex and
ultimately depends on the infrared properties of
QCD. Its investigation requires some nonperturbative approach.

In so doing we start with a tree level
Hamiltonian which after decoupling of the
$c$-quark has the form
$$
H^{tr}_{\Delta S=1}
= {G_F\over\sqrt 2}V_{ud}V^*_{us}Q^u_2.
\eqno(18)
$$
The quantity of interest is an effective theory
realization of that
part of the operator $Q_2$ which is responsible
for a transfer of the kaon into gluons.
Invoking the results of our previous consideration we
can write down this part in the following form
$$
Q^G_2=g^Gf^2_\pi (U^\dagger \chi)_{23}
tr(\partial_\mu U^\dagger\partial^\mu U)
\eqno(19)
$$
where $g^G$ is a dimensionless parameter. We should note that the
chiral representation of the whole operator $Q_2$ contains a large
number of structures but we are interested only in the part
corresponding to the transition with the gluons forming an
intermediate state. Thus the problem is reduced to computing the
chiral coupling constant $g^G$ that can be done by studying the
appropriate Green's function (GF) via QCD sum rules technique. For the
technical reason working with a two point GF is preferable. In the
given decay channel the pions are born by a gluon cloud therefore the
gluon operator $G^a_{\mu\nu}G^a_{\mu\nu}$ can play the role of an
interpolating operator of the pion pair. Thus, it is natural to choose
GF in the following form
$$
G(p)=\int
\langle 0|T{\alpha_s\over\pi}G^a_{\mu\nu}G^a_{\mu\nu}(x)
Q_2(0)|K^0(q)\rangle e^{ipx}dx|_{q=0}.
\eqno(20)
$$

A remark about the chiral limit
for the kaon in eq.~(20) is necessary. The representation
(19) fixes the correct $O(p^4)$ chiral
behavior of the considered decay amplitude and
does not depend explicitly on the kaon momentum.
Keeping a non-vanishing kaon momentum leads
to a shift of the decay amplitude that
lies beyond the accuracy of the present approach.
Thus we can put $q=0$ in eq. (20) and work with
GF depending on one argument only.

Saturating GF in eq.~(20) with $\pi^+\pi^-$ and
$\pi^0\pi^0$ states (the lowest states with
proper quantum numbers), substituting the $Q_2$ operator
and using the low energy theorem (12) one obtains
the following physical representation
$$
G^{ph}(p)=g^G{32 m^2_K\over \pi^2 b f\pi}p^4
\ln ({-p^2\over\mu^2})+O(p^6).
\eqno(21)
$$
The theoretical side reads after making use of {\it OPE}
$$
G(p)={i\over 2\pi^2}\ln ({-p^2\over\mu^2})
{\alpha_s\over\pi}
\langle 0|m_s\bar s_Rg_sG^a_{\mu\nu}
t^a\sigma_{\mu\nu}d_L|K^0(q)\rangle |_{q=0}+O(p^{-2}).
\eqno(22)
$$
Factor $m_s$ in eq.~(22) provides the correct
chiral property of GF and justifies the
representation (19) for the operator $Q_2$.
By reducing the kaon state, one can
transform eq.~(22) into the following expression
$$
G(p)={1\over 4\pi^2}\ln ({-p^2\over\mu^2})
{\alpha_s\over\pi}
f_Km^2_Km^2_0 .
\eqno(23)
$$
where $m_0$ determines a scale of
nonlocality of the quark condensate,
$\langle \bar q g_s G_{\mu\nu}\sigma_
{\mu\nu}q\rangle =m_0^2\langle \bar qq\rangle,
{}~~~m_0^2(1~GeV)=0.8\pm0.2~GeV^2$ [17].
For extracting information about the chiral
coupling constant $g^G$ we use finite
energy sum rules [18] with the result
$$
g^G={3b\over 128}
{f^2_\pi m^2_0\over s^2_0}{\alpha_s(s_0)\over \pi}.
\eqno(24)
$$

To take into account strong interactions at
short distances the operator $Q^G_2$ in the
effective Hamiltonian has to be multiplied by the
corresponding Wilson coefficient $z_2(s_0)$.
Finally, the new contribution to the theoretical
estimate of the $\Delta I=1/2$ amplitude in terms of
the parameter $g_{1/2}$ takes the form
$$
\Delta g_{1/2}=z_2(s_0){\alpha_s(s_0)\over \pi}
{3b m^2_0m^2_K\over 8 s^2_0}.
\eqno(25)
$$
This result needs some comments:

\noindent
1. This next-to-leading in $1/N_c$ expansion
contribution is missed within the factorization
framework and also within any approach where
quark currents in four-quark operators are
replaced by their mesonic counterparts separately.

\noindent
2. The gluon cloud in the intermediate
state does not form a resonance state and,
therefore, the contribution (25) is not
suppressed by a large scalar meson mass.

\noindent
3. In general, some more complicated scalar
colorless gluon configurations, for example,
$f^{abc}G^a_{\mu\nu}G^b_{\nu\lambda}G^c_{\lambda\mu}$,
could appear as intermediate states
in this channel as well. However the theorem  (12)
shows that the two pion form factor of such
configurations could be of the $O(p^4)$ or higher order
in chiral expansion that leads to
a negligible $O(p^6)$ shift of the
decay amplitude.

The question now is what numerical value for the duality
interval $s_0$ has to be used. Actually, the allowed value
of the duality threshold is quite restricted by
the form of the physical spectrum and by the
requirement of absence of uncontrollable
$\alpha_s$ corrections.
To suppress contributions of higher mass states,
for example, a scalar meson $\sigma (0.9~GeV)$, to the
considered channel one has to take $s_0<(0.9~GeV)^2$.
At the same time the
physical representation (20) is obtained in the
leading order in chiral perturbation theory and
the whole procedure is justified until the
chiral expansion parameter $s_0(8\pi^2f^2_\pi)^{-1}$
remains small [13,19]. On the
other hand at the scale $\mu$ less than $0.8~GeV$
the perturbative $\alpha_s$ corrections to
Wilson coefficients become uncontrollable [2] and
for consistency of the approach one has to set
the low limit of the duality interval to be
$s_0>(0.8~GeV)^2$. The reasonable choice
for the duality interval now reads $s_0=(0.8~GeV)^2$.

Let us estimate the uncertainty of our result.
On the physical side of sum rules the errors
related to higher order terms in chiral
expansion, which have been omitted in eq. (21), are, in
general, unknown. But one can hope
that in the spirit of chiral perturbation theory
they are about $25\%$ [13,19].
On the theoretical side of the sum rules the
errors come from two sources. The first one is
the perturbative part of {\it OPE} (the unit
operator) that is
suppressed by an extra loop factor
$\alpha_s/4\pi\sim 10^{-3}$
and can not lead to a sizable change of our
result. Next nonperturbative corrections due
to operators with higher dimensionality seem to
be more important. They start with the
dimension eight operators which have already been
discussed. Numerical estimates would require knowing
the matrix elements of those operators
between the kaon and the vacuum state which are not
available now. But as a first approximation
the relative weight of these corrections can be
represented by the ratio
$$
{\langle
g_s^2 G^a_{\mu\nu}G^a_{\mu\nu}
\rangle\over s_0m_0^2}
{\Gamma (3)\over \Gamma (5)}\sim 10^{-1}
\eqno(26)
$$
where $\langle {\alpha_s\over\pi}G^a_{\mu\nu}G^a_{\mu\nu}
\rangle\sim(330~MeV)^4$ and
the factor $\Gamma (n)$ comes
from a quark loop with $n-1$ gluon field or
mass insertions. The situation here is quite similar to that of
the analysis of $1/m^2_c$ corrections [12] where
a contribution of dimension eight operators
is suppressed rather numerically than
parametrically. Taking into account the uncertainty
of determination of the parameter
$m^2_0$ we estimate the error bound to be about
$40\%$. Numerically one obtains
$$
\Delta g_{1/2}=0.56\pm 0.22
\eqno(27)
$$
at the point $s_0=\mu^2 =(0.8~GeV)^2,~\Lambda_{QCD}=
300~MeV,~z_2(s_0)=1.49~[9]$.

Thus, the new contribution provides about $15\%$
of the observed amplitude
$(17a)$. At the same time it is comparable
with the leading order result for the decay
amplitude obtained by the naive factorization of
the four-quark operator $Q_2$  when all strong
interaction corrections are neglected
$$
g_{1/2}^{fac}=5/9.
\eqno(28)
$$
This implies the strong violation of
factorization in the $O(p^4)$ order in chiral
expansion that leads to additional
enhancement of the theoretical estimate of the
$K\rightarrow\pi\pi$ decay amplitude with the
isospin transfer $\Delta I=1/2$.

To conclude, in the  present paper we consider
a new $K\rightarrow\pi\pi$ decay channel with the
gluons forming an intermediate state. Both
short-distance (perturbative) and long-distance
(non-perturbative) parts of the corresponding
amplitude have been calculated. The latter being
dominant gives an additional enhancement of the
$K\rightarrow\pi\pi$ decay
with the isospin transfer $\Delta I=1/2$
and provides about $15\%$ of the of the
experimentally observed amplitude.
New contribution is of the $O(p^4)$ order
and is lost within the factorization framework.
It allows us to conclude that there is a sizable
violation of the factorization in the $O(p^4)$
order in chiral expansion.
Taking into account corrections of this order along with the
usually considered $\alpha_s$ corrections
to coefficient functions of Wilson expansion can help
to resolve the $\Delta I=1/2$ problem within the SM.

\vspace{0.5cm}
This work was supported in part by Russian Fund for Fundamental Research
under Contract No. 93-02-14428, by Soros Foundation and by
Japan Society for the Promotion of Science (JSPS).

\end{document}